\documentclass[12pt,preprint]{aastex}

\begin{document}

\title{The Energy Distribution of Gamma-Ray Bursts}
\author{David L. Band}
\affil{X-2, MS B-220, Los Alamos National Laboratory, Los
Alamos, NM  87545}
\email{dband@lanl.gov}

\begin{abstract}
The distribution of the apparent total energy emitted by a
gamma-ray burst reflects not only the distribution of the energy
actually released by the burst engine, but also the distribution
of beaming angles.  Using the observed energy fluences, the
detection thresholds and burst redshifts for three burst samples,
I calculate the best-fit parameters for lognormal and power-law
distributions of the apparent total energy.  Two of the samples
include a small number of bursts with spectroscopic redshifts,
while the third sample has 220 bursts with redshifts determined by
the proposed variability-luminosity correlation.  I find different
sets of parameter values for the three burst samples. The Bayesian
odds ratio cannot distinguish between the two model distribution
functions for the two smaller burst samples with spectroscopic
redshifts, but does favor the lognormal distribution for the
larger sample with variability-derived redshifts.  The data do not
rule out a distribution with a low energy tail which is currently
unobservable.  I find that neglecting the burst detection
threshold biases the fitted distribution to be narrower with a
higher average value than the true distribution; this demonstrates
the importance of determining and reporting the effective
detection threshold for bursts in a sample.
\end{abstract}


\section{Introduction}

The growing number of gamma-ray bursts with redshifts has not only
established that most, if not all, bursts are at cosmological
distances, and that up to $10^{54}$ erg are radiated by a burst,
but permits us to determine important intrinsic physical
distributions.  Since we sample a burst's radiation pattern at
only one point, the observed energy fluence can only be related to
the energy flux emitted in our direction; this energy flux can be
expressed as the total energy the burst would have emitted if it
radiated isotropically.  Here I consider the distribution of this
apparent total energy.

The distribution of the apparent total energy is a convolution of
the distribution of the actual energy emitted and the distribution
of the angle into which the emission is beamed, both quantities of
crucial importance in understanding the physics of the progenitor,
the frequency of burst occurrence, and the impact of a burst on
its environment.  Frail et al. (2001) have recently determined the
beaming angles for a burst sample by modeling the
late-time breaks in the temporal decay of the afterglows;
accounting for the beaming angle provides the actual total energy,
which they found is clustered around $E\sim 5\times10^{50}$ erg.  If the
analysis of Frail et al. is indeed correct (there are competing
models for the evolution of the afterglow) then methods similar to
those I develop here will be necessary to determine properly the
beaming angle and total energy distributions.

This study extends the work of Jimenez, Band \& Piran (2001) which
considered lognormal distributions for the apparent total
gamma-ray energy, the peak gamma-ray luminosity, and the total
X-ray afterglow energy. Jimenez et al. expanded the database of
bursts with spectroscopic redshifts by adding bursts for which a
redshift probability distribution could be inferred from the host
galaxy brightness. Currently the bursts with only redshift
probability distributions do not augment the relevant burst
database sufficiently to warrant their inclusion in my study,
although in the future it may be advantageous to include these
bursts if the determination of spectroscopic redshifts does not
keep pace with the detection of bursts and their host galaxies.

Recently correlations have been proposed between burst properties
and their peak luminosities.  Norris, Marani \& Bonnel (1999)
proposed that more luminous bursts have smaller time lags between
energy channels, while Fenimore \& Ramirez-Ruiz (2001) reported
that the light curves of more luminous bursts were more variable;
Schaefer, Deng \& Band (2001) found that applying both
correlations to the same burst sample resulted in consistent burst
luminosities.  The redshift is determined from the derived
luminosities and the observed peak fluxes.  Thus these
correlations can give us a large burst sample with redshifts from
which the energy distribution can be determined.

In \S 2 I discuss the methodology for finding the best parameter
values for a given functional form of the distribution, and for
comparing different functional forms.  I also evaluate the
sensitivity to small burst samples, and demonstrate the importance
of considering the detection threshold in determining the
distributions.  The energy distributions resulting from two small
burst samples with spectroscopic redshifts and from a large sample
with redshifts from the variability-luminosity correlation are
described in \S 3.  Finally, \S 4 summarizes my conclusions.

\section{Methodology}

\subsection{The Likelihood Function}

I begin with an assumed distribution $p(E \,|\, \vec{a_j}, M_j,
I)$ where $E$ is the apparent total burst energy, $\vec{a_j}$ is
the set of parameters which characterize the $j$th model
distribution function represented by $M_j$, and $I$ specifies
general assumptions about the distribution function.  I use
$p(a\,|\,b)$ to mean the probability of $a$ given $b$.  Below I
present the distribution functions used in this study. A
distribution is assumed to be universal and not a function of
redshift (i.e., no evolution), or of properties of the host
galaxy, burst, etc.  Of course, once a sufficiently large sample
is available, the energy distributions for burst subsets can be
investigated.  I use normalized distributions since the burst rate
is not of interest here. The observed energy fluence $F$ and the
burst energy $E$ are related by $F=E (1+z)/4\pi D_L(z)^2 = E C(z)$
where $D_L(z)$ is the luminosity distance.  It is in calculating
the luminosity distance that the burst redshift and a cosmological
model are required.  Here I assume H$_0$=65 km s$^{-1}$
Mpc$^{-1}$, $\Omega_m=0.3$ and $\Omega_\Lambda=0.7$.

This energy distribution is converted into $p(F \,|\, \vec{a_j},z,
M_j,I)$, the probability of obtaining the energy fluence $F$ given
the parameters $\vec{a_j}$ for model $M_j$, the burst redshift
$z$, and other assumptions $I$ (e.g., the choice of cosmological
model). However, the observed fluences are not drawn from this
probability distribution but from $p(F \,|\, F_T,\vec{a_j},
z,M_j,I)$, the normalized distribution which is truncated below
$F_T$, the minimum fluence at which that particular burst would
have been included in the burst sample.

I now form the likelihood
\begin{equation}
   L_j=\prod_{i=1}^N p(F_i \,|\, F_{T,i},\vec{a_j},z_i,M_j,I)
\end{equation}
where the $i$th burst has energy fluence $F_i$, threshold fluence
$F_{T,i}$, and redshift $z_i$.  In the ``frequentist'' framework
best fit parameters are typically found by maximizing $L_j$.

The likelihood is $L_j=p(D \,|\, \vec{a_j},M_j,I)$, where $D$ is
the set of observed fluences and fluence thresholds.  The Bayesian
method of estimating the parameters is to calculate
\begin{eqnarray}
\langle \vec{a_j} \rangle &=& \int d\vec{a_j} \,\vec{a_j}\,
   p(\vec{a_j}\,|\, D,M_j,I) \\
   &=& {{\int d\vec{a_j} \,\vec{a_j}\, p(D\,|\, \vec{a_j},M_j,I)
   p(\vec{a_j}\,|\, M_j,I)} \over
   {\int d\vec{a_j} \, p(D\,|\, \vec{a_j},M_j,I)
   p(\vec{a_j}\,|\, M_j,I)}}
   = {{\int d\vec{a_j} \,\vec{a_j}\,
   \prod_{i=1}^N p(F_i \,|\, F_{T,i},\vec{a_j},z_i,M_j,I)
   p(\vec{a_j}\,|\, M_j,I)} \over{\int d\vec{a_j} \,
   \prod_{i=1}^N p(F_i \,|\, F_{T,i},\vec{a_j},z_i,M_j,I)
   p(\vec{a_j}\,|\, M_j,I)}} \nonumber
\end{eqnarray}
where $p(\vec{a_j}\,|\, M_j,I)$ is the prior for the parameters
$\vec{a_j}$.  If $\Lambda_j = p(D\,|\, \vec{a_j},M_j,I)
p(\vec{a_j}\,|\, M_j,I)$ is sharply peaked, then the expectation
value of the parameters occurs at the peak of $\Lambda_j$.  Note
that $\Lambda_j$ is the likelihood in eq.~1 times the priors for
the parameters. In some sense the priors indicate the natural
variables in terms of which the priors are constant. For example,
here the burst energies vary over a number of decades, and thus I
assume that the priors for average or cutoff energies are
logarithmic.  Consequently maximizing $\Lambda_j$ is equivalent to
maximizing the likelihood in terms of the logarithm of the energy
parameters.  This is the methodology I use here; I do not attempt
to integrate the integrals in eq.~2.

For each burst the cumulative probability is
\begin{equation}
P(F_i \,|\, F_{T,i},\vec{a_j},z_i,M_j,I) =
   \int_{F_i}^\infty p(F \,|\, F_{T,i},\vec{a_j},z_i,M_j,I)\, dF
   \quad .
\end{equation}
If the assumed energy distribution function is an acceptable
characterization of the observations (which would be the case if
the model $M_j$ is correct) and all the assumptions are valid
(e.g., the cosmological model is correct), then the cumulative
probabilities $P(F_i)$ should be uniformly distributed between 0
and 1, and have an average value of $\langle P(F_i) \rangle=
1/2\pm (12N)^{-1/2}$ for $N$ bursts in the sample.

\subsection{Model Comparison}

As mentioned above, the average of the cumulative probability
$\langle P(F_i) \rangle$ should be 1/2 (within a quantifiable
uncertainty) if the distribution function describes the
observations satisfactorily.  Thus the value of this statistic for
model $M_j$ is a measure of the acceptability of that model.
Further, a comparison of the values for different models is a
measure of the relative merits of these models; of course, this
comparison should account for the expected uncertainty in the
value of this statistic.

The Bayesian framework provides a clear prescription for comparing
models through the odds ratio.  Let $p(M_j \,|\, D , I)$ be the
posterior probability for the $j$th model $M_j$ given the data
$D$. Then the odds ratio comparing the $j$th and $k$th models is
$O_{jk} = p(M_j \,|\, D , I)/ p(M_k \,|\, D , I)$. But by Bayes'
Theorem $p(M_j \,|\, D , I)\propto p(M_j \,|\, I) p(D \,|\, M_j ,
I)$ where the normalizing factor is independent of $M_j$, and thus
cancels in forming the odds ratio. I assume that no model is
favored a priori; therefore the ``priors'' $p(M_j \,|\, I)$ are
the same for all $M_j$, and cancel in forming the odds ratio. Here
$D$ consists of the observed energy fluences (and the detection
thresholds). Thus $p(M_j \,|\, D , I)\propto \prod_i p(F_i \,|\,
F_{T,i},z_i,M_j,I)$. However, we begin with the more fundamental
probabilities $p(F_i \,|\, F_{T,i},z_i,M_j,\vec{a_j}, I)$, which
are functions of the model parameters $\vec{a_j}$. In determining
the preferred model we are uninterested in the specific model
parameters, and therefore we ``marginalize'' over the parameters
$\vec{a_j}$: $ p(F_i \,|\, F_{T,i},z_i,M_j,I) = \int d\vec{a_j} \,
p(\vec{a_j}\,|\,M_j,I) \prod_i p(F_i \,|\, F_{T,i},z_i, M_j,
\vec{a_j}, I)$, where $p(\vec{a_j}\,|\,M_j,I)$ is the prior for
the parameters of the $j$th model.  Thus
\begin{equation}
O_{jk} = {{\int d\vec{a_j} \, \prod^N_i
   p(F_i \,|\, F_{T,i},z_i,M_j,\vec{a_j}, I)
   p(\vec{a_j}\,|\,M_j,I)} \over
   {\int d\vec{a_k} \, \prod^N_i
   p(F_i \,|\, F_{T,i},z_i,M_k,\vec{a_k}, I)
   p(\vec{a_k}\,|\,M_k,I) }} \quad .
\end{equation}
It will be noted that the odds ratio is the ratio of the
likelihoods for each model (eq.~1) marginalized over the model
parameters. Often the prior for the parameters is set equal to a
delta function at the parameter values which maximize the
likelihood, which is of course circular reasoning.  In this common
formulation, the odds ratio is the ratio of the peak values of the
likelihoods for each model.  Below I present priors defined in
terms of the expected parameter ranges, in which case the
integrals in eq.~4 must be integrated. I present values of the
odds ratio with both the delta function priors and the more
refined priors.  It is possible that with the delta function
priors the odds ratio may favor one model, but with the more
refined priors another model is favored, as is indeed the case
here.

\subsection{Distribution Functions}
\subsubsection{Lognormal Distribution}

I assume that $E$ has a log-normal distribution
\begin{equation}
p(E \,|\, E_0 ,\sigma) d(\ln E) =
   {1\over{\sqrt{2\pi}\sigma}}
   \exp \left[-{{(\ln E_0 -\ln E)^2}
   \over{2\sigma^2}} \right] d(\ln E) \quad .
\end{equation}
Thus the fluence $F$ also has a log-normal distribution.  Note
that $\sigma$ is a width in logarithmic space, and the linear
change of variables from $E$ to $F$ does not affect this width. As
discussed above, we need to consider the range over which the
fluence could actually have been observed, i.e., for fluences
above the threshold $F_T$. The resulting normalized fluence
probability distribution is
\begin{equation}
p_{\hbox{obs}}(F \,|\, F_T, E_0 ,\sigma,z) d (\ln F) =
   {{{1\over{\sqrt{2\pi}\sigma}}
   \exp \left[-{{\left(\ln\left[ E_0 C(z)\right]-\ln F \right)^2}
   \over{2\sigma^2}}\right] \theta(F-F_T) d(\ln F)}
   \over{{1\over 2}
   \left(1+\hbox{erf}\left[ {{\ln\left[ E_0
   C(z)\right]-\ln(F_T) }\over {\sqrt{2}\sigma}} \right]\right)}}
\end{equation}
where $\theta(x)$ is the Heaviside function (1 above $x=0$, and 0
below) and $C(z)=(1+z)/4\pi D_L(z)^2$ converts energies in the
burst's frame to fluences in our frame; the denominator results
from integrating over the numerator from $F_T$ to infinity.

For the Bayesian formulation we also need the priors for the model
parameters.  There is no reason to favor one energy over another
over many energy decades, and thus I assume the prior is constant
in logarithimic space:  $p(E_0) dE_0 = (E_0 \ln [E_u/E_l])^{-1}
dE_0=(\ln[E_u/E_l])^{-1} d\ln E_0= (\log_{10} [E_u/E_l])^{-1}
d\log_{10} E_0$ where $E_u$ and $E_l$ are the upper and lower
limits of the permitted range (because of the logarithmic
dependence, the result is not very sensitive to the precise
values).  I will use $E_l = 10^{51}$ erg and $E_u = 10^{54}$ erg.
Similarly, I have no a priori information about $\sigma$, and
therefore assign it a uniform prior between 0 and 5. Note that the
distribution function is explicitly a function of $\ln E_0$ and
$\sigma$, and the priors indicate that these are indeed the
natural parameters.

\subsubsection{Single Component Power Law Distribution}

I assume that the underlying energy probability distribution is
\begin{eqnarray}
p(E\,|\,E_1, E_2, \alpha) dE =&
   {{\left(1-\alpha\right)E_2^{\alpha-1}}
   \over{1-(E_1/E_2)^{1-\alpha}}} E^{-\alpha} dE \quad
   & ; \quad E_1\le E \le E_2,\quad \alpha \ne 1 \nonumber \\
=& {{E^{-1}}\over{\ln(E_2/E_1)}} dE & ; \quad E_1\le E \le E_2,\quad
\alpha = 1 \quad .
\end{eqnarray}
If $E_1$ is extended to 0 or $E_2$ to infinity, then we must
restrict $\alpha$ to be less than or greater than 1, respectively.
The expected fluence probability distribution, accounting for the
fluence threshold $F_T$, is
\begin{eqnarray}
p(F\,|\,F_T,E_1, E_2, \alpha) dF =&
   {{\left(1-\alpha\right) [E_2 C(z)]^{\alpha-1}} \theta(F-F_T)
   \over{1-(\max [E_1,F_T/C(z)]/E_2)^{1-\alpha}}} F^{-\alpha} dF
   & ; \quad E_1\le F/C(z) \le E_2,\quad \alpha \ne 1 \nonumber \\
=& {{F^{-1}\theta(F-F_T)}\over{\ln(E_2/\max [E_1,F_T/C(z)])}} dF &
   ; \quad E_1\le F/C(z) \le E_2, \quad \alpha = 1
\end{eqnarray}
where again $C(z)=(1+z)/4\pi D_L(z)^2$ converts burst energies to
fluences.

As with $E_0$ for the lognormal distribution, I have no reason to
prefer any value of the energy limits $E_1$ and $E_2$ over many
energy decades, and therefore I again use logarithmic priors for
these two energies.  For definiteness, I will assume that $E_1$
can have a value between $10^{49}$ and $10^{52}$ erg, and $E_2$
between $10^{52}$ and $10^{55}$ erg.  The spectral index is
assumed to have a uniform prior between $-2.5$ and 2.5.

\subsection{Data}

The methodology discussed above requires the energy fluence $F$,
the threshold fluence $F_T$ and the redshift $z$ for each burst. I
consider 3 samples.  The B9 sample consists of 9 bursts with BATSE
data and spectroscopic redshifts. The fluences were calculated by
fitting the BATSE spectrum accumulated over the entire burst with
the ``GRB'' function (Band et al. 1993), and then integrating the
resulting fit over the 20--2000~keV energy range in the burst's
rest frame and over the time during which the spectrum was
accumulated. The resulting fits are presented in Jimenez et al.
(2001).  There are some bursts for which the high energy power law
in the GRB function has an index $\beta<-2$ (where $N\propto
E^\beta$) and thus the fluence depends crucially on the high
energy cutoff (or roll-off) which must exist for a finite fluence
but which could not be determined from the BATSE data. The
spectroscopic redshifts are taken from Frail et al. (2001).

The limiting fluence $F_T$ is more difficult to determine. The
bursts in our sample must have been intense enough to be first
detected and then properly localized.  In addition, a decision was
made to attempt to observe the afterglow and thus determine the
redshift. These threshold quantities have generally not been
reported.  Note also that detectors almost never trigger on the
fluence but usually trigger on the peak count rate sampled over a
time bin of a specified duration.  Here I will assume that the
ratio of the observed to threshold fluence, $F/F_T$, is the same
as the ratio of the observed to threshold peak count rate, $C_{\rm
max}/C_{\rm min}$, for the BATSE data.  These thresholds are most
likely underestimates of the true thresholds.  The $C_{\rm
max}/C_{\rm min}$ ratio is a standard part of the BATSE catalog
(BATSE team 2001); however, in some cases the BATSE team did not
calculate this quantity because of data gaps, in which case I
estimated this ratio from the lightcurves.  This sample is
described by Table~1.

The C17 sample consists of the 17 bursts with spectroscopic
redshifts and fluences in Frail et al. (2001).  This sample is
basically a superset of the B9 sample with an additional BATSE
burst for which there are no spectra, and bursts observed by {\it
Beppo-SAX} and {\it Ulysses}.  For those bursts without reported
detection thresholds I use a fluence threshold of
$10^{-6}$~erg~cm$^{-2}$.

Finally, the F220 sample uses the 220 bursts in Fenimore \&
Ramirez-Ruiz (2001) with redshifts determined by the
variability-luminosity correlation.  This sample was selected to
have a peak count rate accumulated on the 256 millisecond
timescale of greater than 1.5~cts~s$^{-1}$, which provides a
well-defined detection threshold.  The fluences were taken from
the BATSE catalog without any k-corrections.  Note that Bloom et
al. (2001) find that the k-correction for the 20--2000~keV energy
range is of order unity.

It should be noted that for the first two samples the redshifts
are reliable but the detection threshold is very uncertain.  Even
when the detection threshold is known for the gamma-ray portion of
the burst, the effective threshold for optical follow-up
observations has not been reported.  On the other hand, the
detection threshold for the third sample is known, but the
validity of the variability-luminosity correlation is still not
well established, and the uncertainty in the resulting redshifts
is not considered.

\subsection{Simulations}

To determine the sensitivity of the methodology to the number of
bursts and to demonstrate the importance of considering the
fluence threshold, I ran a series of simulations.  For each
simulation I first created between 100 and 500 simulated databases
to which I then applied the methodology described above to
determine the parameters of their energy distribution.  For some
simulations I found the parameters with the fluence thresholds
used in creating the database or with much smaller thresholds.

For each simulated burst I needed a redshift, a burst energy and a
fluence threshold.  The redshifts were drawn from a distribution
which is similar to the proposed cosmic star formation rate.  The
fluence threshold was drawn from a uniform logarithmic
distribution over one decade.  The burst energy was drawn from a
lognormal distribution with specified central energy and
logarithmic width, as long as the resulting fluence was greater
than the fluence threshold.

Figure 1 shows the importance of considering the fluence threshold
in calculating the likelihood.  As can be seen, the best fitted
parameter values cluster around the input central energy
$E_0=10^{53}$~erg and logarithmic width $\sigma=1.5$ when the
fluence threshold is considered (asterisks), but cluster around a
higher energy and narrower width when the fluence threshold is
neglected (diamonds). The fluence threshold removes low energy
bursts, resulting in a narrower apparent distribution which is
shifted to higher energy. Each of the 100 simulated datasets had
80 bursts, and a fluence threshold between $10^{-6}$ and $10^{-5}$
erg cm$^{-2}$.  Figure~2 shows that the likelihood contours for a
sample analyzed with the correct fluence thresholds (left panel)
and thresholds a factor of 10 smaller (right panel) differ
significantly.

Table~2 gives the width of the distributions of the parameters of
the lognormal distribution, $\log E_0$ and $\sigma$, for databases
with 9, 20, 40 and 80 bursts.  As expected, the distributions
become narrower as the number of bursts increases. As can be seen,
a database with 40 bursts should give satisfactory best-fit
parameter values.

\section{Results}

As can be seen from Table~3, the parameter values at the
likelihood maximum for the two distribution functions differ for
the three burst samples, and the 90\% confidence ranges from one
burst sample do not always include the parameters from the other
samples.  Note that the lower energy cutoff $E_1$ for the simple
power law distribution is not fitted.  The fits are insensitive to
$E_1$ values smaller than the lowest energy for which any burst in
the sample would have been detected, and $E_1$ cannot be greater
than the smallest observed burst energy; the difference between
these two limits is very small. Figures~3--8 show the likelihood
contours for the fitted parameters.  A ridge of somewhat lower
likelihood values curves towards lower values of $E_0$ and higher
values of $\sigma$ for the lognormal distributions. Thus the data
do not strongly exclude a broader distribution which includes
lower energy bursts which are not detected because their fluence
is below the threshold.  Consequently, the 90\% confidence bounds
for the parameters are quite broad.  For the power law
distributions, the favored high energy cutoff is at the highest
observed energy in the burst sample; if the bursts truely have a
power law distribution, this cutoff is most likely somewhat
greater.

Table~3 also provides the average of the cumulative probabilities
for each sample and distribution.  As can be seen, these averages
are within $\sim 1\sigma$ of the expected value of 1/2, indicating
that the distributions are acceptable descriptions of the data,
and that this statistic cannot be used to discriminate between
distributions.  The actual distributions of these cumulative
probabilities are also presented by Figures~3--8.

Finally, Table~3 presents the likelihood ratios and the Bayesian
odds ratios comparing the lognormal to power law distributions.
The likelihood ratio is a Bayesian odds ratio using delta function
priors set to the parameters which maximize the likelihoods (which
violates the definition of a prior). The Bayesian odds ratios
given here use the priors described in \S 2.3. The ratios show
that using a broad prior distribution favors the lognormal
distribution over the power law distribution.  Based on the odds
ratio, the B9 and C17 samples are insufficient to discriminate
between the two distribution functions.  On the other hand, the
odds ratio favors the lognormal distribution for the F220 sample.

\section{Discussion and Conclusions}

The parameters and parameter ranges differ for the three different
burst samples.  It is not clear whether we yet have a sufficiently
large, properly defined burst sample from which to calculate the
energy distribution.  The two samples with spectroscopic redshifts
do not have correct detection thresholds: the threshold for
detecting the burst itself is usually reported, but the intensity
threshold which triggers further localization and spectroscopic
redshift determination has not been reported. Indeed, there may
not yet be a formal definition of such a followup threshold.  Thus
these two samples are flawed.  On the other hand, the validity of
the variability-determined redshifts has not yet been proven,
although the detection threshold was defined in choosing the burst
sample. The importance of the detection thresholds for statistical
studies of burst samples argues for well-defined (and reported)
intensity thresholds for triggering the followup observations of
the expected large number of {\it HETE-II} and {\it SWIFT} burst
localizations.

The distributions of cumulative probabilities and the averages of
these distributions indicate that the two functional forms used
here are sufficient to describe the distribution of energies;
consequently, I concluded that the data do not justify trying more
complicated distribution functions at this time.  The Bayesian
odds ratio does not distinguish between these two functional forms
for the two samples with spectroscopic redshifts, although it does
favor the lognormal distribution for the large F220 sample with
redshifts derived from the variability of the burst lightcurves.

The energy distribution cannot be determined for energies below
the lowest energy threshold (i.e., the lowest burst energy
corresponding to the fluence thresholds).  Indeed, Hakkila et al.
(1996) make a distinction between the ``observed'' and
``intrinsic'' luminosity functions in their study of luminosity
functions for cosmological bursts based on the shape of the peak
flux distribution; they point out that the observed distribution
may be much narrower than the intrinsic distribution. In my study
the true low energy cutoff of the power law distribution cannot be
determined. Similarly, the likelihood contours for the lognormal
distribution do not rule out broader distributions with lower
central energies.

We anticipate that the {\it HETE-II} and {\it SWIFT} missions will
result in the construction of a large burst sample with
spectroscopic redshifts.  Properly defined subsets can be studied
to identify trends with burst redshift, duration, and other
properties.  As we move from the study of individual bursts to the
study of burst ensembles, we must define and report the criteria
(e.g., detection thresholds) by which the burst samples are
collected.
\acknowledgements
This work was performed under the auspices of
the U.S. Department of Energy by the Los Alamos National
Laboratory under Contract No. W-7405-Eng-36.

\clearpage

\begin{figure}
\plotone{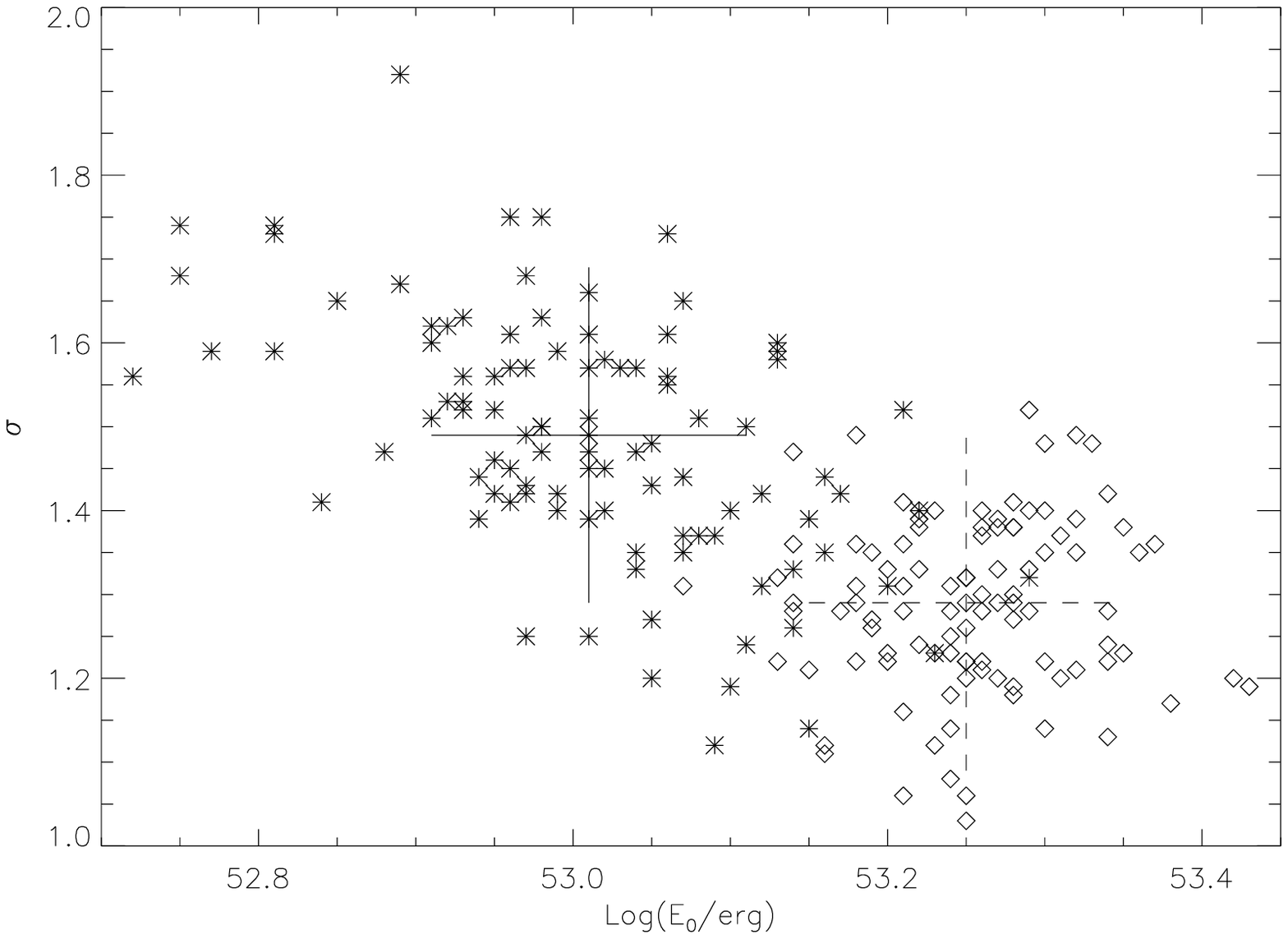}
\caption{Best-fit parameter values including (asterisks) or
neglecting (diamonds) the fluence threshold in calculating the
likelihood for 100 simulated databases with 80 bursts each. The
model lognormal energy distribution had a central energy of
$E_0=10^{53}$~erg and a logarithmic width of $\sigma=1.5$. The
median values for the fits including (large solid cross) or
neglecting (large dashed cross) the fluence threshold are
indicated.  The bursts were drawn from a redshift distribution
similar to that of star formation, and the fluence threshold was
between $10^{-6}$ and $10^{-5}$ erg~cm$^{-2}$.}
\end{figure}

%\begin{figure}
%\caption{Distributions of the best-fit central energy $E$
%(panel~a) and logarithmic width (panel~b) for simulations with 9
%(solid histogram), 20 (dashed), 40 (dot-dashed), and 80
%(3~dots-dashed) bursts in a simulated database. A lognormal energy
%distribution with no fluence thresholds was used; a redshift
%distribution which peaks at $z\sim1.5$ was assumed.  The
%histograms have been normalized so that the bins centered on the
%input values of $E_0=100$ (in units of $10^{51}$ erg) and
%$\sigma=1.5$ have unit heights.  The histogram boundaries and
%heights have been shifted by small amounts for greater
%legibility.}
%\end{figure}

\clearpage

\begin{figure}
\plotone{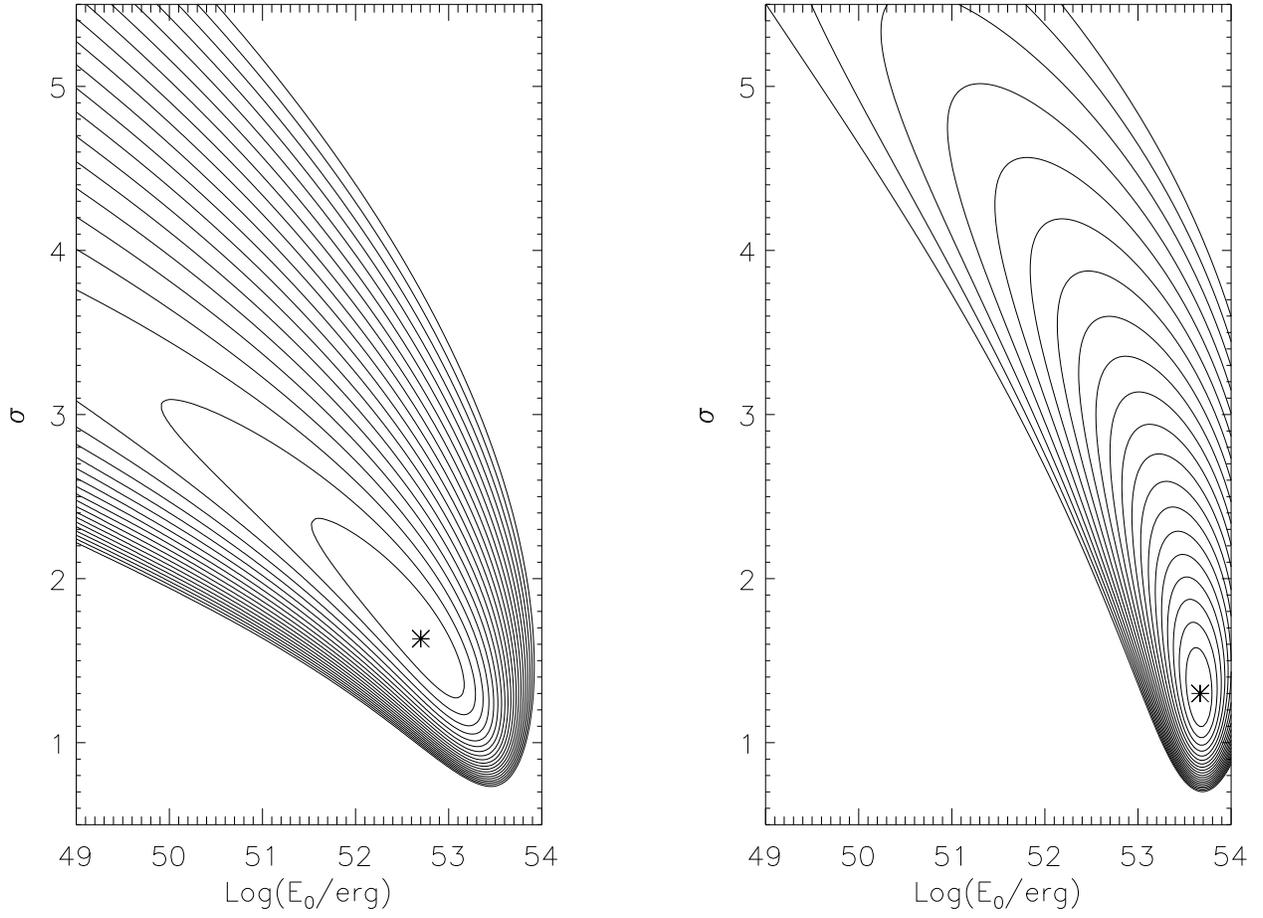}
\caption{Comparison of the likelihood contours for a sample
analyzed with the correct fluence thresholds (left) and thresholds
a factor of 10 smaller (right).  The sample of 9 bursts was drawn
from a lognormal distribution with $E_0 = 10^{53}$~erg and
$\sigma=1.5$, a redshift distribution similar to star formation,
and a uniform fluence threshold between $10^{-5}$ and $10^{-4}$
erg~cm$^{-2}$.}
\end{figure}

\clearpage

\begin{figure}
\plotone{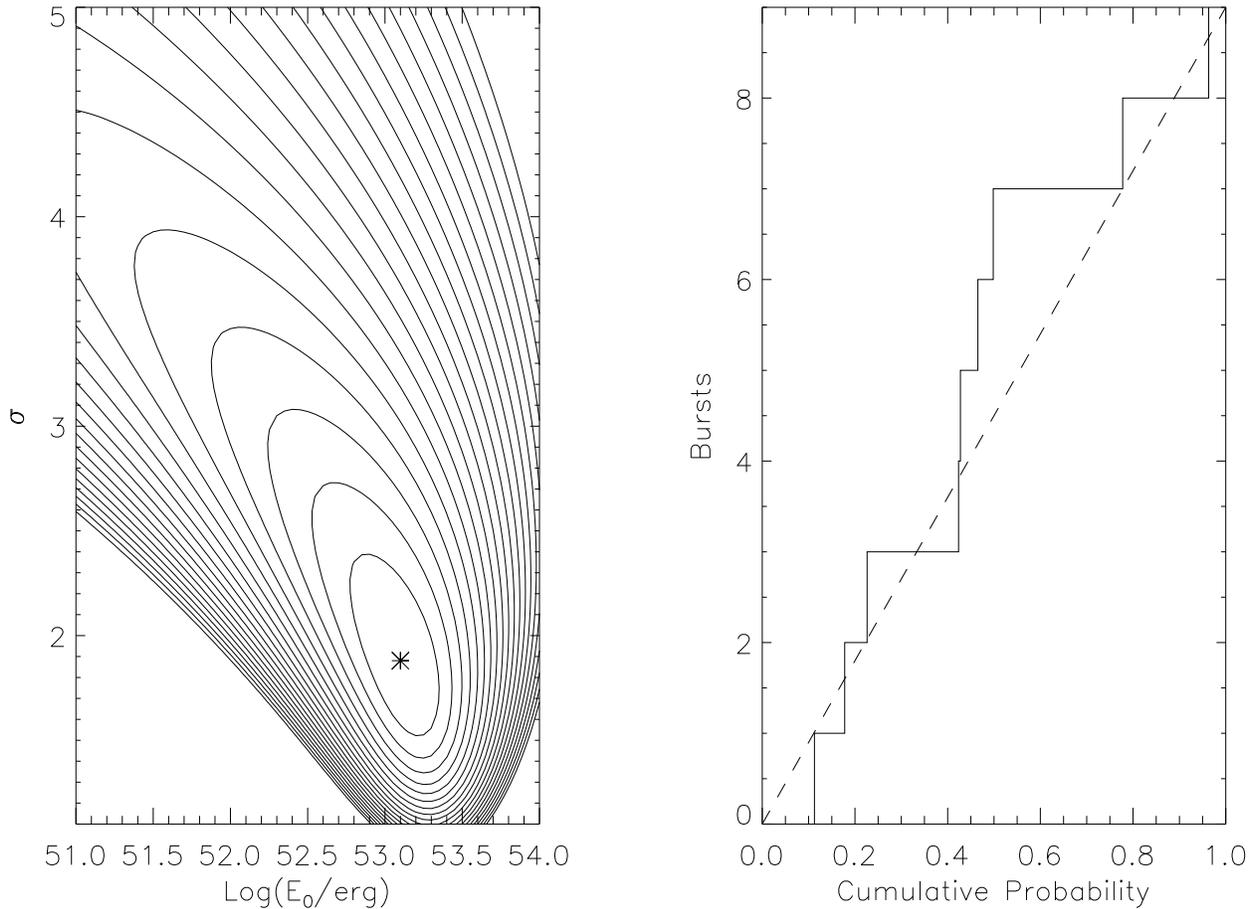}
\caption{Contour plot of the likelihood for the lognormal energy
distribution for the B9 sample (left panel).  The parameters are
the central energy $E_0$ and the logarithmic width $\sigma$.  The
asterisk indicates the location of the maximum likelihood while
contours are spaced by $\Delta \log_{10} L=0.1$ starting from the
maximum value. Cumulative distribution of the cumulative
probability for each burst assuming their energies are drawn from
the best-fit lognormal energy distribution for the B9 sample
(right panel).  The dashed line is the expected distribution.}
\end{figure}

\clearpage

\begin{figure}
\plotone{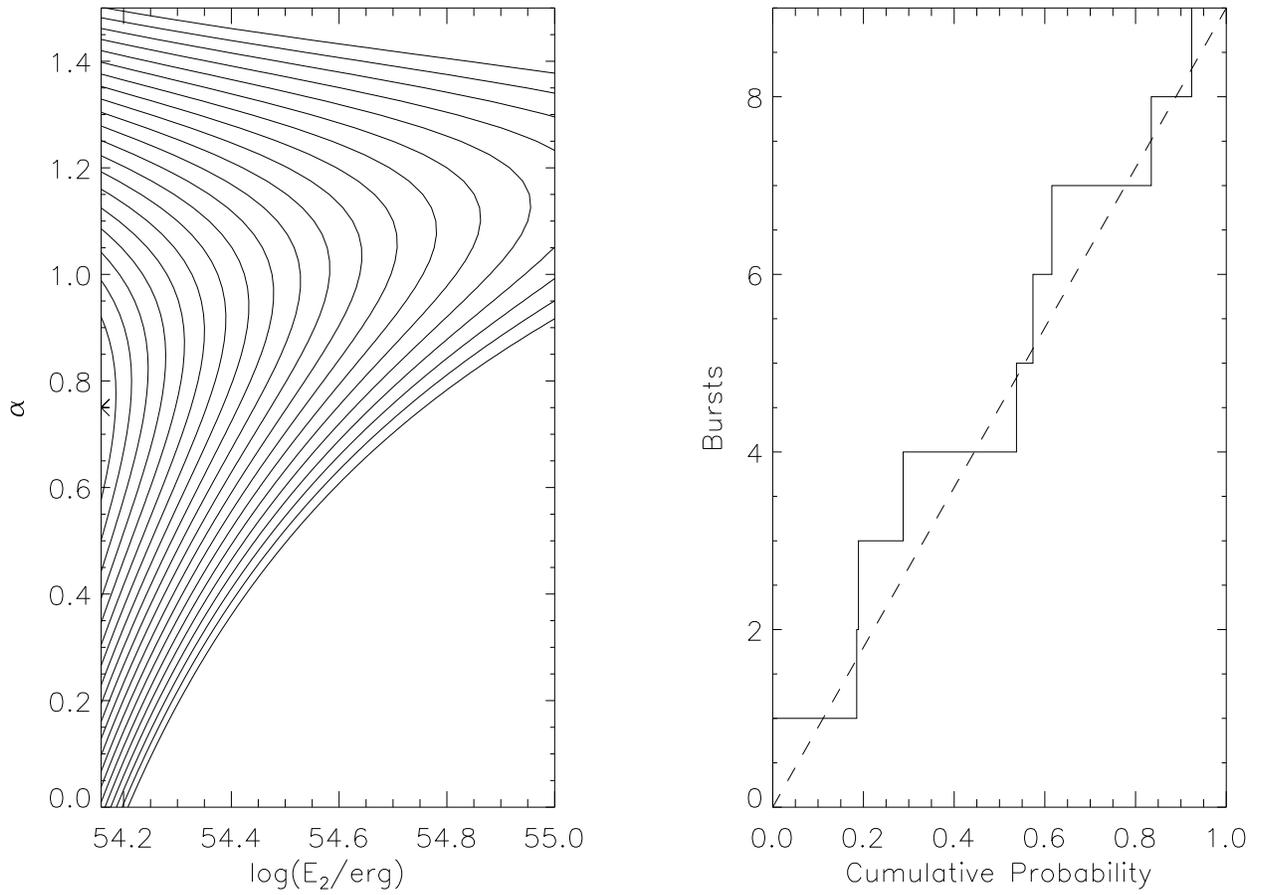}
\caption{The same as Figure~3, but for a simple power law energy
distribution for the B9 sample.  The parameters are the upper
cutoff energy $E_2$ and the power law index $\alpha$.}
\end{figure}

\clearpage

\begin{figure}
\plotone{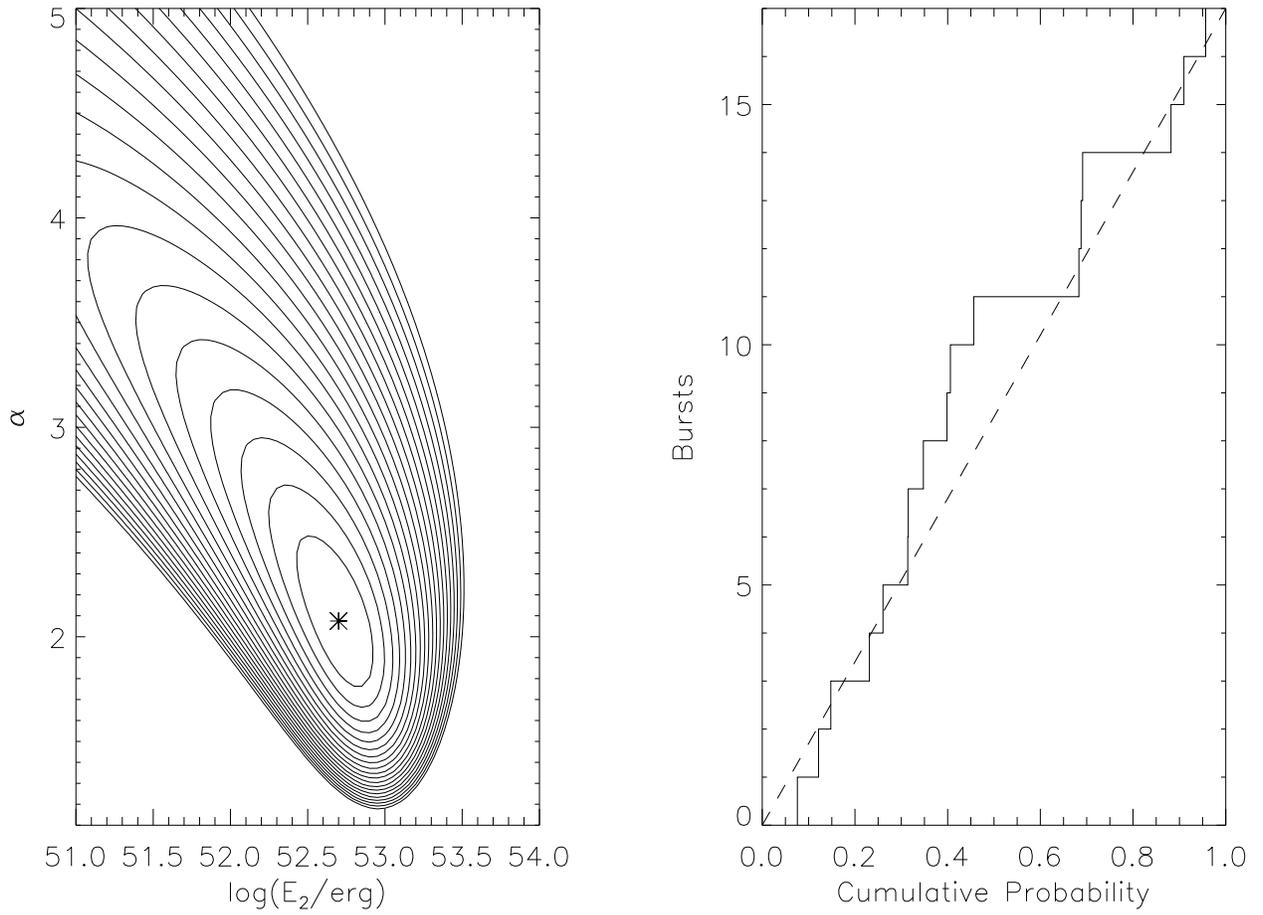}
\caption{The same as Figure 3 for the lognormal distribution and
the C17 sample.}
\end{figure}

\clearpage

\begin{figure}
\plotone{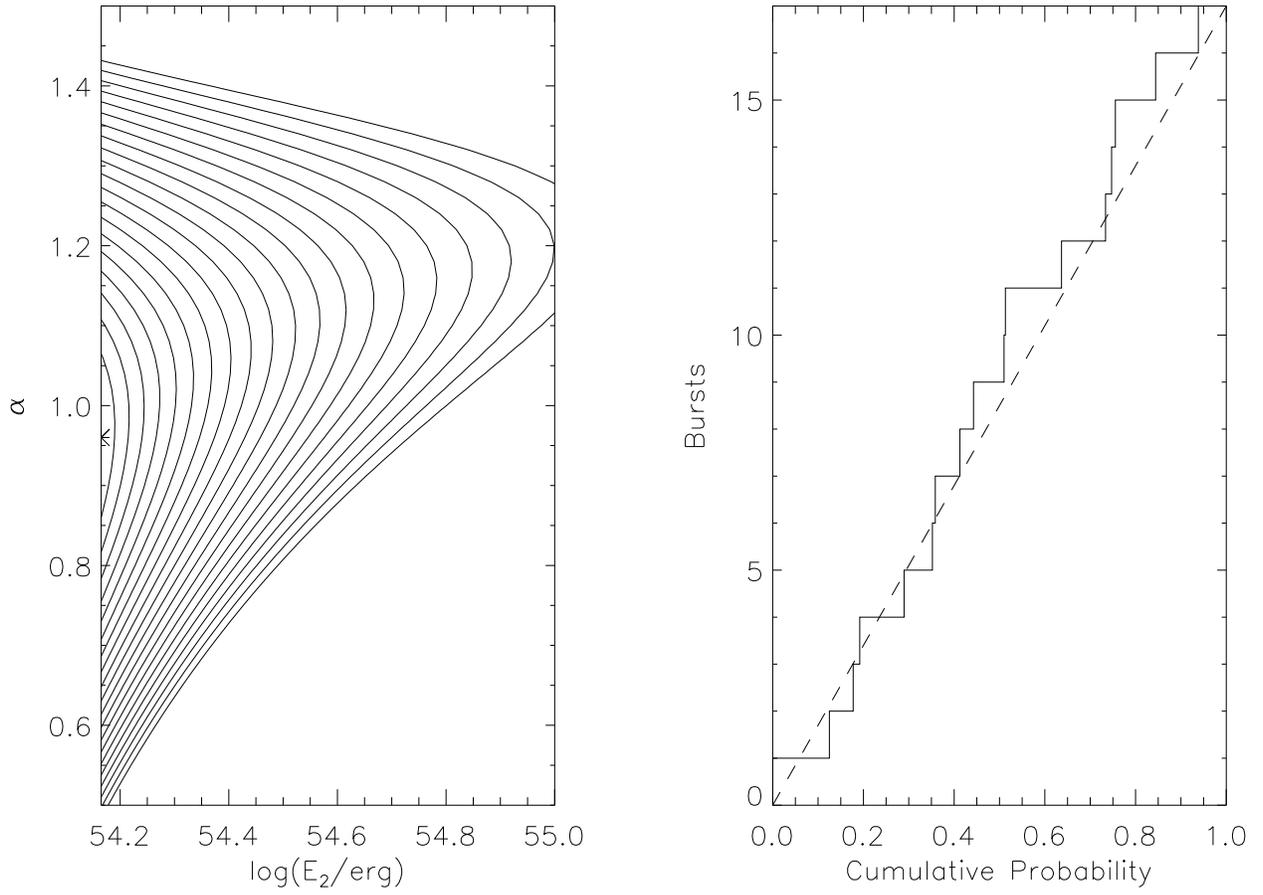}
\caption{The same as Figure 4 for the power law distribution and
the C17 sample.}
\end{figure}

\clearpage

\begin{figure}
\plotone{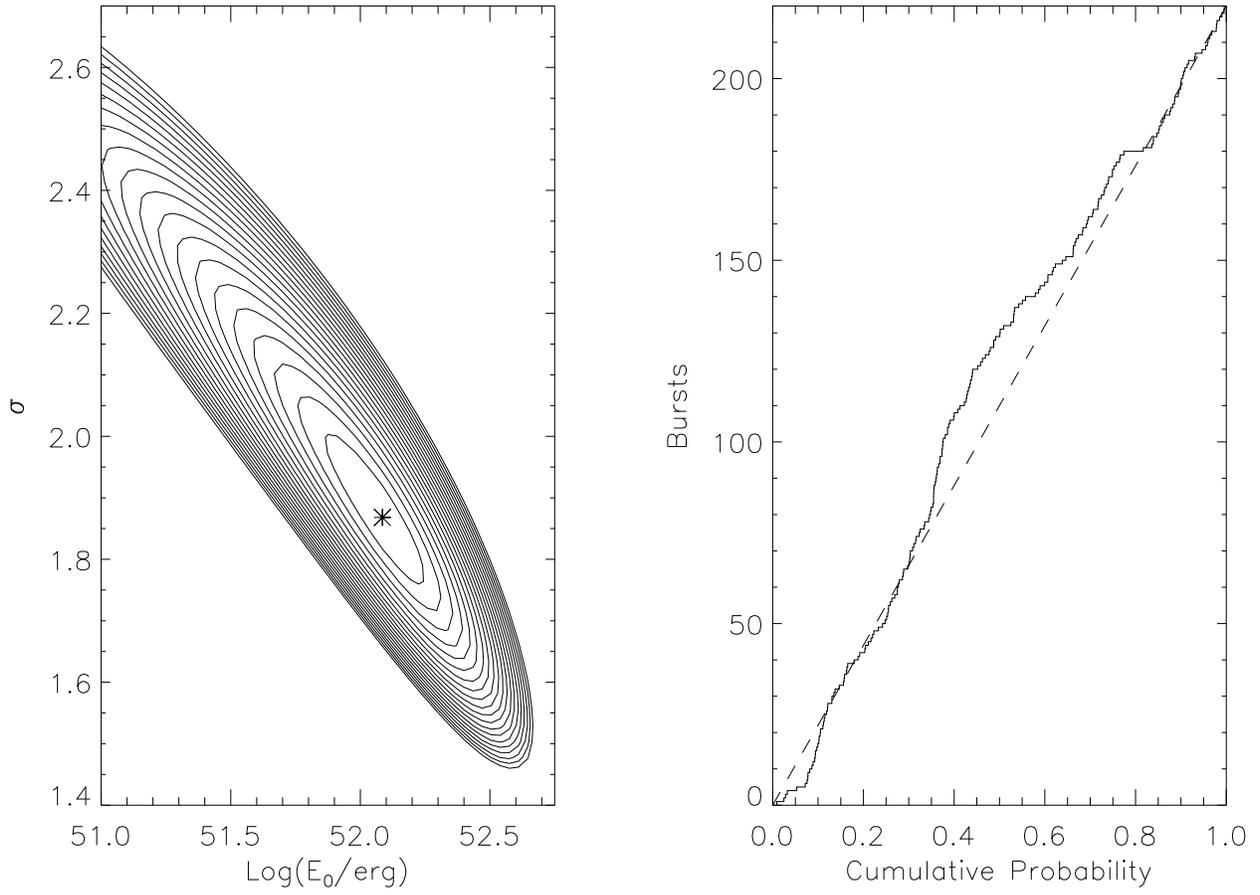}
\caption{The same as Figures 3 and 5 for the lognormal distribution
and the F220 sample.}
\end{figure}

\clearpage

\begin{figure}
\plotone{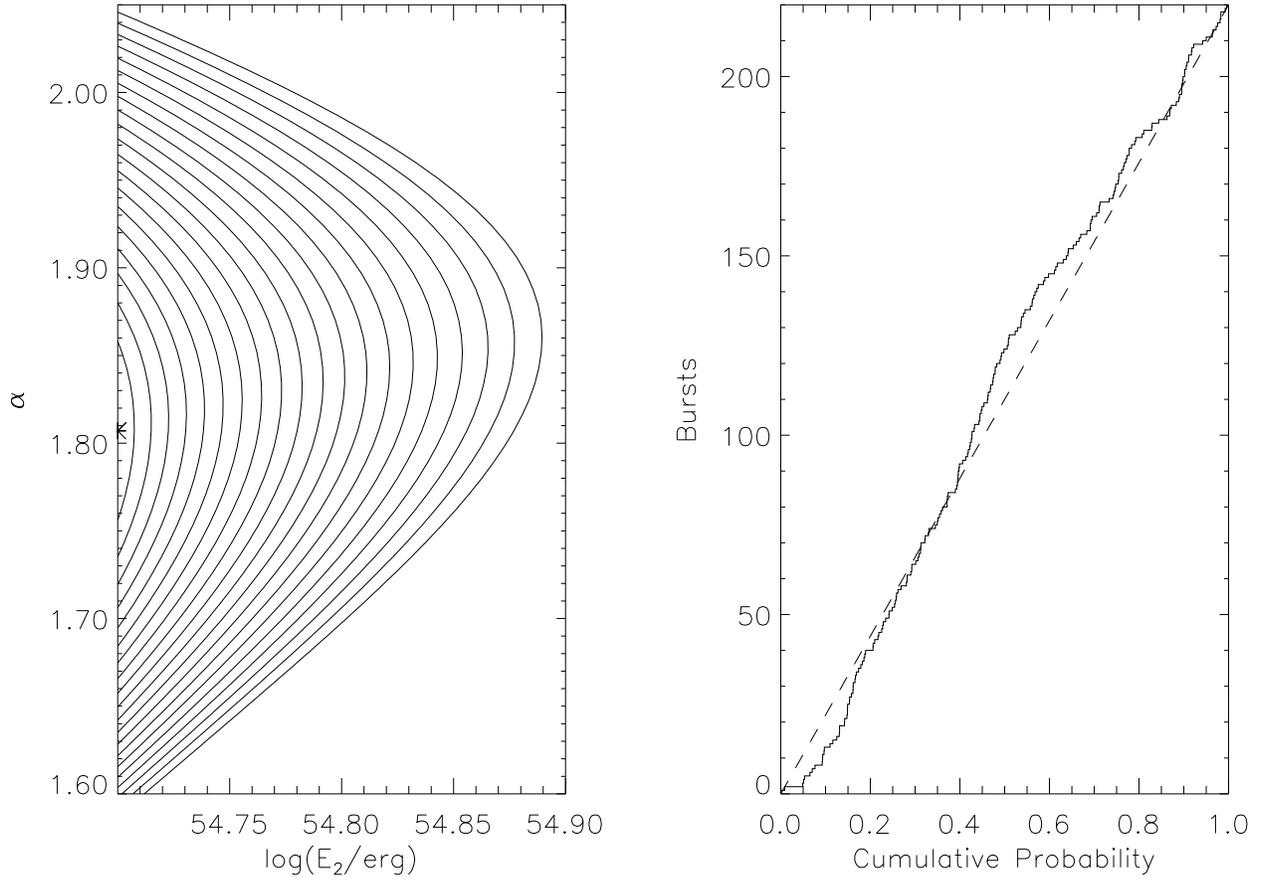}
\caption{The same as Figures 4 and 6 for the power law
distribution and the F220 sample.}
\end{figure}

\clearpage

\begin{deluxetable}{l l l l c c c}
\tablecolumns{7}
\tablewidth{0pt}
\tablecaption{\label{Table}The BATSE Gamma-Ray Bursts Sample.}
\tablehead{
\colhead{Burst} &
\colhead{$F_{\rm obs}$\tablenotemark{a}} &
\colhead{$F_{\rm burst}$\tablenotemark{b}} &
\colhead{$z_{\rm obs}$} &
\colhead{${{C_{\rm max}}\over{C_{\rm min}}}$} &
\colhead{$E_{\rm obs}$\tablenotemark{c}} &
\colhead{$E_{\rm burst}$\tablenotemark{d}} \\
& \colhead{(erg cm$^{-2}$)}
& \colhead{(erg cm$^{-2}$)}
&&&
\colhead{($10^{51}$ erg)}
& \colhead{($10^{51}$ erg)}}
\startdata
%
% z--Metzger et al. 1997a, IAU 6676
% R=25.55--Zharikov, S. V., Sokolov, V. V., and Baryshev, Yu. V. 1998, GCN 31
% $1.73\times 10^{-3}$ & $-1.191$ & $-1.831$ &480.84&
970508 & $3.18\times 10^{-6}$ & $2.59\times 10^{-6}$ &
0.835& 3.3\tablenotemark{f} & 6.734 & 5.482 \\
%
% z--Djorgovski talk
%
% $1.15\times 10^{-2}$ & $-0.704$ & $-2.072$ & 229.74&
970828 & $9.57\times 10^{-5}$ & $7.88\times 10^{-5}$ &
0.958& 20\tablenotemark{f} & 267.1 & 219.3 \\
%
% z--Kulkarni et al. 1998a, Nature, 393, 35
% R=26.5 Odewahn et al. 1998, ApJ, 509, L5
% $7.23\times 10^{-3}$ & $-0.783$ & $-2.574$ &155.96&
971214 & $9.44\times 10^{-6}$ & $7.59\times 10^{-6}$ &
3.412& 7.32\tablenotemark{e} & 261.3 & 210.7 \\
%
% Kulkarni et al. 1998b, Nature, 395, 663
% $4.70\times 10^{-3}$ & $-1.266$ & \nodata &161.20&
%980425& $3.87\times 10^{-6}$ & 0.0085& 2.5\tablenotemark{d} & \\
%
% z Djorgovski et al. ApJL 508, 17, 1998
% R=22.8 Bloom, J., & Kulkarni, S. 2000, GCN702
% $4.41\times 10^{-3}$ & $-1.314$ & $-2.396$ &370.26&
980703& $2.26\times 10^{-5}$ & $2.13\times 10^{-5}$ &
0.966& 3.08\tablenotemark{e} & 63.94 & 60.18 \\
%
% z Bloom et al. 1999, ApJ, 518
% R=23.7 Fruchter et al. 1999, ApJ, 516, 689
% $2.62\times 10^{-2}$ & $-0.900$ & $-2.476$ &549.51&
990123& $2.68\times 10^{-4}$ & $1.93\times 10^{-4}$ &
1.600& 80.1\tablenotemark{e} & 1996. & 1438. \\
%
% $1.51\times 10^{-2}$ & $-1.370$ & $-2.152$ &449.78&
990506 & $1.94\times 10^{-4}$ & $1.69\times 10^{-4}$ &
1.2 & 50\tablenotemark{f} & 838.8 & 854.0 \\
%
% $7.96\times 10^{-3}$ & $-1.275$ & $-2.670$ &174.24&
990510 & $2.26\times 10^{-5}$ & $2.32\times 10^{-5}$ &
1.619 & 19.3\tablenotemark{e} & 172.0 & 176.8 \\
%
% $1.29\times 10^{-1}$ & $-1.234$ & $-2.184$ &414.83&
991216 & $1.93\times 10^{-4}$ & $1.70\times 10^{-4}$ &
1.02 & 144.\tablenotemark{e} & 611.1 & 534.1 \\
%
% $1.68\times 10^{-2}$ & $-0.688$ & $-2.068$ & 98.982 &
000131 & $4.18\times 10^{-5}$ & $2.71\times 10^{-5}$ & 4.5
& 3\tablenotemark{f} & 1791. & 1159. \\
\enddata

\tablenotetext{a}{Fluence over 20--2000~keV in the observer's frame.}
\tablenotetext{b}{Fluence over 20--2000~keV in the burst's frame.}
\tablenotetext{c}{Gamma-ray energy over 20--2000~keV in the observer's frame,
assuming isotropic emission, $H_0=65$
km s$^{-1}$ Mpc$^{-1}$, $\Omega_M$=0.3 and $\Omega_\Lambda$=0.7.}
\tablenotetext{d}{Gamma-ray energy over 20--2000~keV in the burst's frame.}
\tablenotetext{e}{From the online BATSE catalog.}
\tablenotetext{f}{Estimated from lightcurve.}

\end{deluxetable}

\clearpage

\begin{deluxetable}{l c c}
\tablecolumns{3}
\tablewidth{0pt}
\tablecaption{\label{Table2}Width of Parameter Distribution}
\tablehead{
\colhead{Number of Bursts} &
\colhead{Width\tablenotemark{a}$\quad$of $\log E_0$} &
\colhead{Width\tablenotemark{a}$\quad$of $\sigma$} \\
\colhead{in Sample} &
\colhead{Distribution} &
\colhead{Distribution}}
\startdata
9 & 0.38 & 0.47 \\
20 & 0.40 & 0.52 \\
40 & 0.20 & 0.31 \\
80 & 0.13 & 0.19 \\
\enddata
\tablecomments{In these simulations 100 samples were constructed with the
indicated number of bursts per sample.  The burst energies were drawn from
a lognormal distribution with central energy $E_0=10^{53}$~erg and logarithmic
width $\sigma=1.5$.  The redshift distribution mimics the cosmic star formation
rate, and the threshold fluence was between $10^{-6}$ and $10^{-5}$ erg cm$^{-2}$.
The best fit parameters were found by maximizing the likelihood.}
\tablenotetext{a}{The width given is the range within which 1/2 the simulated
bursts fell.}

\end{deluxetable}

\clearpage

\begin{deluxetable}{l c c c}
\tablecolumns{4}
\tabletypesize{\footnotesize}
%\tablewidth{6.in}
\tablecaption{\label{Table3}Comparison of Model Distributions}
\tablehead{
\colhead{Quantity} &
\colhead{B9\tablenotemark{a}} &
\colhead{C17\tablenotemark{b}} &
\colhead{F220\tablenotemark{c}} }
\startdata
$E_0$\tablenotemark{d} & 125.9 & 52.48 & 11.79 \\
&1.6--315 & 1.6--100 & 2.--23.4 \\
$\sigma$\tablenotemark{e} & 1.87 & 2.06 & 1.88 \\
& 1.5--4.5 & 1.7--4.15 & 1.65--2.3 \\
$E_1$\tablenotemark{f} & 1.6 & 0.55 & 0.12 \\
$E_2$\tablenotemark{g} & 1440 & 1460 & 5000 \\
& 1440--3550 & 1460--3350 & 5000--6100 \\
$\alpha$\tablenotemark{h} & 0.74 & 0.96 & 1.81 \\
& 0.4--1.2 & 0.75--1.25 & 1.7--1.94 \\
$\langle p\rangle_{ln}$\tablenotemark{i} & 0.4525 & 0.4638 &
0.4753 \\
$\langle p\rangle_{pl}$\tablenotemark{j} & 0.4608 & 0.4723 &
0.4892 \\
$\sigma_{\langle p\rangle}$\tablenotemark{k} & 0.0962 & 0.0700 &
0.0195 \\
Likelihood Ratio\tablenotemark{l} \qquad \qquad  &  \qquad \qquad$4.29\times10^{-2}$  \qquad\qquad&
 \qquad\qquad$5.61\times10^{-2}$\qquad \qquad & \qquad\qquad $4.38\times10^{2}$  \qquad \qquad\\
Odds Ratio\tablenotemark{m} & 1.19 & 1.65 & $9.92\times10^{3}$ \\
\enddata
\tablenotetext{a}{Sample of 9 BATSE bursts with spectroscopic
redshifts and fitted spectra (Table~1).}
\tablenotetext{b}{Sample
of 17 bursts with spectroscopic redshifts (Frail et al. 2001).}
\tablenotetext{c}{Sample of 220 bursts with redshifts derived from
variability redshifts (Fenimore \& Ramirez-Ruiz 2001).}
\tablenotetext{d}{The central energy of the lognormal
distribution, in units of $10^{51}$~erg.  The following line in
the table provides the 90\% confidence range.}
\tablenotetext{e}{The logarithmic width (in units of the energy's
natural logarithm) for the lognormal distribution.  The following
line provides the 90\% confidence range.}
\tablenotetext{f}{The low energy cutoff of the power law
distribution, in units of $10^{51}$~erg.  This energy has been set
to the lowest threshold energy for the sample.}
\tablenotetext{g}{The high energy cutoff of the power law
distribution, in units of $10^{51}$~erg.  The following line
provides the 90\% confidence range.}
\tablenotetext{h}{The power law index of the power law
distribution, $p(E)\propto E^{-\alpha}$.  The following line
provides the 90\% confidence range.}
\tablenotetext{i}{Average of the cumulative probabilities for the
lognormal distribution; 1/2 is expected.}
\tablenotetext{j}{Average of the cumulative
probabilities for the power law distribution; 1/2 is
expected.}
\tablenotetext{k}{The
standard deviation $1/[12N]^{1/2}$ of the average of the
cumulative probabilities for a sample of $N$ bursts.}
\tablenotetext{l}{Ratio of the maximum likelihood for the
lognormal to power law distributions.  A value greater than 1
favors the lognormal distribution.}
\tablenotetext{m}{Odds ratio
comparing the lognormal to power law distributions.  A value
greater than 1 favors the lognormal distribution.}

\end{deluxetable}

\end{document}